\documentclass[aps,pre,reprint,floatfix]{revtex4-2}

\usepackage{amsmath}
\usepackage{amssymb}
\usepackage{graphicx}
\usepackage{bm}
\usepackage[separate-uncertainty]{siunitx}
\usepackage{orcidlink}

\graphicspath{{figures/}}

\renewcommand{\vec}[1]{\bm{#1}}
\renewcommand{\tensor}[1]{\overline{\overline{#1}}}
\newcommand{\order}[1]{\mathcal{O}\!\left\{ #1 \right\}}
\newcommand{\abs}[1]{\left\vert #1 \right\vert}
\newcommand{\real}[1]{\Re\left\{ #1 \right\}}

\begin{document}

\title{Emergent activity powers a path to chaos for pairs of acoustically trapped spheres}

\author{Matthew K. Gronert\,\orcidlink{0000-0002-6552-3027}}
\author{Ben Cao\,\orcidlink{0009-0002-2389-6746}}
\author{Ella M. King\,\orcidlink{0000-0002-6374-3819}}
\altaffiliation{Department of Chemical and Biological Engineering,
  Northwestern University, Evanston, Illinois 60208, USA}
\author{David G. Grier\,\orcidlink{0000-0002-4382-5139}}
\affiliation{Department of Physics and Center for Soft Matter Research,
  New York University, New York, New York 10003, USA}

\date{\today}

\begin{abstract}
  Pairs of spheres sharing an acoustic trap sometimes spin intermittently despite having no self-propulsion, no obvious power source, and no symmetry-breaking mechanism, switching between rocking and spinning within a single trajectory. Nonreciprocal wave scattering powers this emergent activity, and drag shapes their dynamics by coupling rotations to translations. Mismatched pairs can pass through a supercritical Hopf bifurcation into a quasiperiodic torus on which the pair's orbit reverses regularly and, farther from equilibrium, into deterministic chaos. Experiments motivate this mechanism and share its qualitative signatures; simulations reveal a full noise-free route from passivity to chaos, and suggest how analogous dynamics can emerge in other active-matter systems.
\end{abstract}

\maketitle

\begin{figure}
    \centering
    \includegraphics[width=\columnwidth]{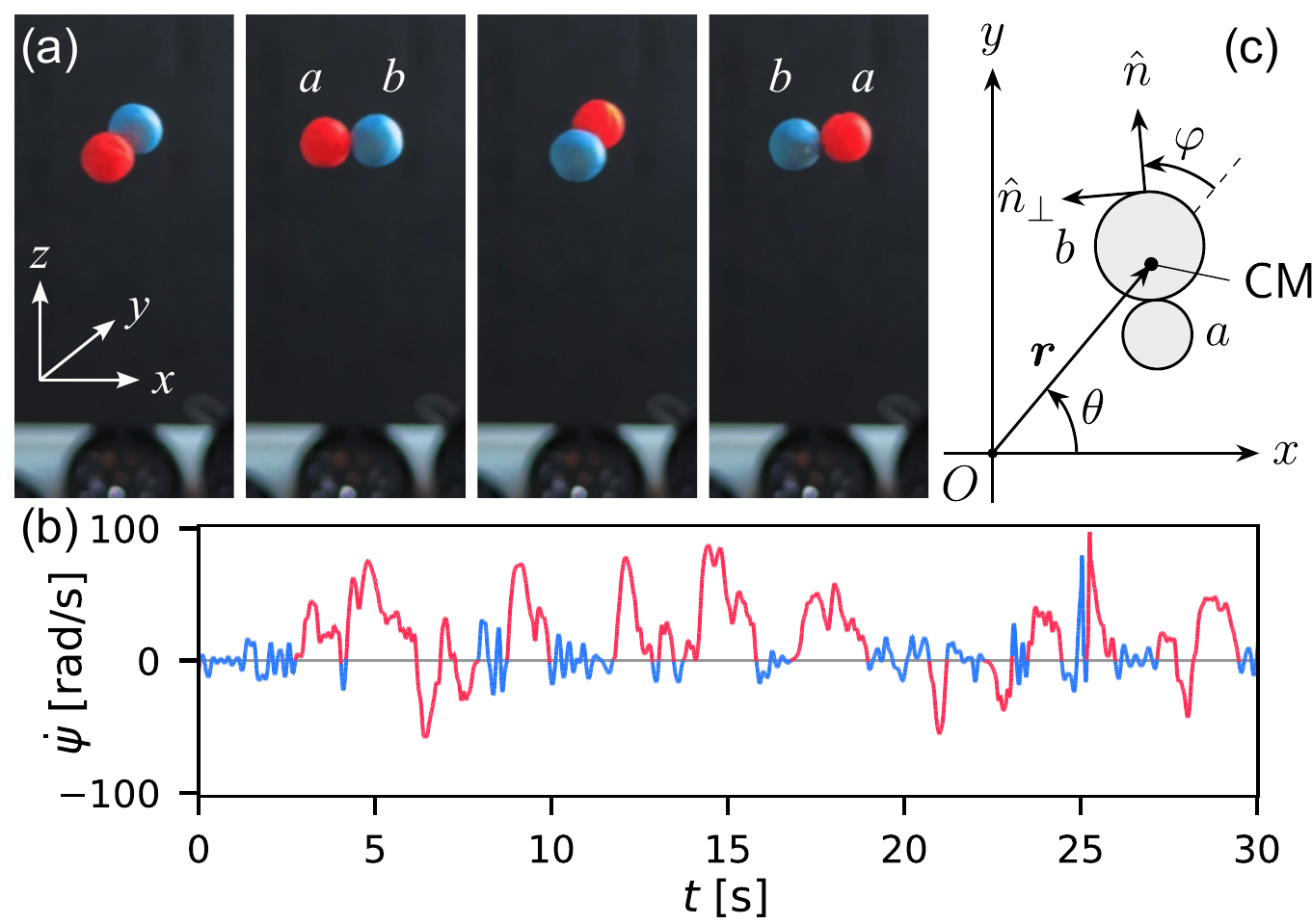}
    \caption{(a) Video sequence \cite{supplementalvideo1} at \qty{67}{\ms} intervals
    of a pair of millimeter-scale
    beads, $a$ and $b$, spontaneously spinning around
    the $\hat{z}$ axis of
    an acoustic levitator.
    (b) Measured rotation rate, $\dot\psi(t) = \dot{\varphi}(t) +
    \dot{\theta}(t)$, colored by dynamical state
    (red: spinning, at least one full rotation,
    blue: rocking) for another pair of beads with $a = \qty{0.895(5)}{\mm}$
    and $b = \qty{0.930(5)}{\mm}$.
    (c) Geometry:
    The origin, $O$, is centered on the trap.
    The beads' center of mass (CM) is displaced from $O$
    by $\vec{r}$ at angle $\theta$ relative to $\hat{x}$.
    The pair's axis, $\hat{n}$, points from $a$ to
    $b$ and makes angle $\varphi$ with respect to $\hat{r}$.
    The transverse unit vector is $\hat{n}_\perp = \hat{z} \times \hat{n}$.
    }
    \label{fig:schematic}%
\end{figure}

A small sphere levitated in mid-air by a standing acoustic wave
normally settles into static equilibrium at a node.
Introducing a second sphere into the same node
can cause surprising departures from quiescence.
Some pairs twitch.
Others, such as the example in Fig.~\ref{fig:schematic}(a), spin vigorously,
spontaneously breaking symmetry while overcoming
viscous drag.
Such spinning is never steady, but rather proceeds
episodically with frequent pauses and reversals.
We show that these behaviors are manifestations of \emph{emergent activity} \cite{king2025scattered}:
the spheres power their own autonomous motion
by harvesting energy from the sound wave, without the continuous
energy consumption that defines conventional active matter
\cite{fodor2016far,henkes2011active,palacci2013living,dauchot2019dynamics,fruchart2021nonreciprocal}.
Nonreciprocal wave-mediated interactions have been shown to
propel small collections of spheres deterministically,
inducing steady translations and rotations
\cite{king2025scattered,yifat2018reactive},
and sustained oscillations
\cite{morrell2026nonreciprocal}.
Intermittent spinning embodies emergent dynamics of a richer kind.

A minimal model inspired by experimental
observations explains both twitching and
intermittent spinning.
Nonreciprocal interactions push
an otherwise quiescent pair away from equilibrium,
and viscous drag converts that drive into a torque that creates both librational and spinning orbits.
Neither kind of orbit is stable, however.
Instead, a levitated pair
either settles onto
a quasiperiodic torus on which its
orbital direction reverses regularly, or else
descends into deterministic chaos.

Other models for nonsteady active dynamics
build in self-propulsion
\cite{fodor2016far,henkes2011active,palacci2013living}
and rely on noise to escape stable limit cycles
\cite{dauchot2019dynamics,baconnier2025selfaligning}.
Neither is needed for the emergent phenomena described here.
We propose that analogous inherent paths from passivity to chaos
can arise in a general class of systems
whose degrees of freedom are coupled by both
reciprocal and nonreciprocal forces.
Emergent microscopic chaos could then supply such systems with
intrinsic fluctuations that organize their collective behavior,
without invoking external noise sources.

The video sequence in Fig.~\ref{fig:schematic}(a)
shows a typical experimental realization of spontaneous
spinning by
a pair of expanded polystyrene (EPS) beads trapped
in the central node of a TinyLev2 ultrasonic levitator
\cite{marzo2017tinylev} operating at \qty{40}{\kilo\hertz}.
The axis of the standing wave is aligned with $\hat{z}$,
and the selected video frames capture the beads
spinning around $-\hat{z}$ at \qty{3.6(1)}{\hertz}.
Relevant physical quantities are reported in the Supplemental Material
(Sec.~\ref{app:parameters}).
The full video \cite{supplementalvideo1} shows that the beads stop spinning after a few seconds and then
resume several seconds later in the opposite direction.
Intermittent spinning continues indefinitely, as shown in the
measured rotation rate of another pair in
Fig.~\ref{fig:schematic}(b).

The minimal model depicted in Fig.~\ref{fig:schematic}(c)
consists of two solid spherical beads
levitated in an acoustic standing wave
and subject to viscous drag.
The beads have radii $a$ and $b$,
with $b \geq a$, and are assumed for simplicity to be composed of the same material.
The standing wave confines the beads along $\hat{z}$
so that their motion is restricted to the
horizontal $(x, y)$ plane.
The beads are held in contact by the trap
and by the spheres'
wave-mediated K\"onig interaction \cite{konig1891hydrodynamisch,hoffmann1996visualization,silva2014acoustic,king2025scattered},
which tends to be attractive at small separations.
The system's instantaneous configuration
therefore can be described with three degrees of freedom:
two, $\vec{r} = (r, \theta)$, for the displacement
of the center of mass (CM) from the center of the trap,
and one for the pair's orientation, $\varphi$,
relative to $\hat{r}$.
Defining the pair's absolute orientation to be $\psi = \theta + \varphi$ means that
$\hat{n} = (\cos\psi, \sin\psi)$ is the unit vector pointing from
bead $a$ to bead $b$.
The perpendicular 
unit vector is $\hat{n}_\perp = \hat{z} \times \hat{n}$.

In this coordinate system,
the beads' centers are located at $\vec{r}_a = \vec{r} - \ell_a\hat{n}$ and
$\vec{r}_b = \vec{r} + \ell_b\hat{n}$,
where $\ell_a$ and $\ell_b$
are the distances from the CM
to the beads' centers.
The beads' moment of inertia about their CM is
$I = M \ell_a \ell_b$, where
$M = \tfrac{4}{3} \pi \rho_p (a^3 + b^3)$ is their
total mass given their mass density, $\rho_p$.

The ultrasonic levitator operates at a high enough frequency
that a sphere of radius $a$
experiences a pressure node as a
time-averaged potential energy well
\cite{gorkov1962forces,bruus2012acoustofluidics,settnes2012forces,zang2022natural,morrell2023acoustodynamic}
whose curvature,
\begin{equation}
  \kappa_a = 3 f k^2 \frac{p_0^2}{\rho_m c_m^2} \, a^3,
  \label{eq:stiffness}
\end{equation}
is proportional to the sphere's volume.
The force scale is set by
the amplitude of the pressure wave, $p_0$,
and depends on the medium's density and sound speed,
$\rho_m$ and $c_m$, respectively.
The wave number, $k$, establishes a characteristic
length scale for the system.
The factor $f \in (0, 1]$ in Eq.~\eqref{eq:stiffness}, the trap's aspect ratio, is the ratio of its in-plane stiffness
to its axial stiffness.
An ideal planar standing wave exerts no in-plane
restoring force, and therefore has $f = 0$.
The type of levitator used for this study
focuses sound waves into the trapping volume,
yielding $f \in \numrange{0.06}{0.2}$ depending on details \cite{marzo2017tinylev,marrara2024optical,drewitt2024mightylev,argyri2024customized}.

Because both spheres are trapped in the same node and are made of the same material,
they have the same stiffness-to-mass ratio.
The net confining potential
for the pair therefore reduces to
\begin{equation}
  U_\text{trap}(r)
  =
  \tfrac{1}{2} \kappa_a \, r_a^2
  +
  \tfrac{1}{2} \kappa_b \, r_b^2
  =
  \tfrac{1}{2} \kappa \, r^2,
  \label{eq:confining_potential}
\end{equation}
where we have omitted an additive constant.
The trap's effective stiffness, $\kappa = \kappa_a + \kappa_b$,
establishes a characteristic frequency scale
for the system,
\begin{equation}
    \omega_0
    =
    \sqrt{\frac{9}{4\pi} \frac{1}{\rho_p \rho_m}}
    \, \frac{k p_0}{c_m} ,
    \label{eq:omega0}
\end{equation}
that is independent of the beads' radii.
The natural frequency for beads' in-plane oscillations
is $\sqrt{f} \omega_0$.
Because $U_\text{trap}(r)$
does not depend on $\hat{n}$, however, it exerts no torque about the CM and thus does not account for the beads' tendency to spin.

\begin{figure}
    \centering
    \includegraphics[width=\columnwidth]{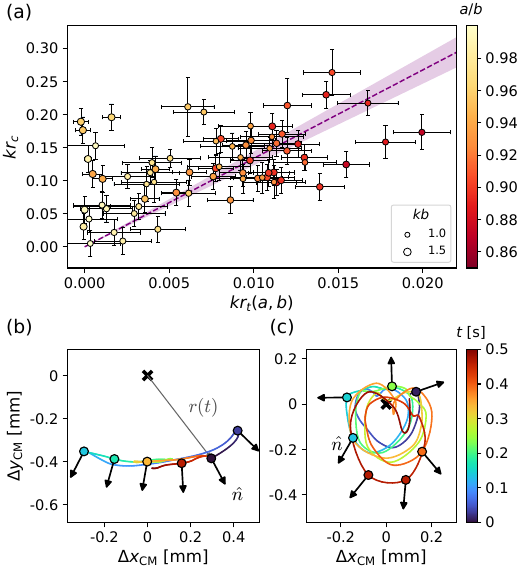}
    \caption{(a) Static equilibrium: measured CM displacement, $r_c$,
    of pairs of EPS beads versus predictions,
    $r_t(a,b)$, for a trap with $f = 1$.
    The diameter of each plot symbol is proportional to $kb$; color indicates $a/b$.
    Linear regression yields
    $f = \num{0.075(6)}$ for the trap's aspect ratio.
    (b) Libration: measured CM trajectory, $\vec{r}(t)$, relative to the trap ($\times$),
    of a pair of beads actively rocking:
    $a = \qty{1.242(5)}{\mm}$, $b = \qty{1.988(3)}{\mm}$.
    (c) Spinning: $a = \qty{1.317(5)}{\mm}$, $b = \qty{1.529(5)}{\mm}$.
    Color in (b) and (c) indicates elapsed time,
    circles mark representative data points, and arrows show
    the measured pair orientation, $\hat{n}$, at those times.}
    \label{fig:experiment}
\end{figure}

Levitated beads typically are smaller
than the wavelength of sound, which
means that we can adopt the Rayleigh approximation, $ka \leq kb < 1$,
in formulating their wave-mediated interactions.
To leading order in the beads'
radii, the force on
sphere $a$ due to waves scattered
by sphere $b$ is central, attractive and reciprocal
\cite{king2025scattered}:
\begin{subequations}
\begin{equation}
    \vec{F}_0
    =
    \frac{\pi}{2} \frac{p_0^2}{k^2 \rho_m c_m^2}
    (ka)^3 (kb)^3 \, \Phi(ka + kb) \, \hat{n} .
\end{equation}
Its magnitude
depends on the contact separation through
\cite{weiser1984interparticle,silva2014acoustic,king2025scattered}
\begin{equation}
    \Phi(x) = \frac{\cos(x) + x \sin(x)}{x^2} .
\end{equation}
\label{eq:F0}%
\end{subequations}
The derivation of Eq.~\eqref{eq:F0} assumes the
pressure wave to be uniform,
which is reasonable for $f \ll 1$.

Size-dependent corrections
\cite{king2025scattered}
modify the force on sphere $a$
due to sphere $b$ by a multiplicative factor
that breaks symmetry under exchange
\cite{morrell2026nonreciprocal,king2025scattered}:
\begin{equation}
  \vec{F}_{ab}
  =
  \vec{F}_0 \left[
    1 - \tfrac{3}{10}(ka)^2 - \tfrac{19}{18}(kb)^2\right] .
  \label{eq:Fab}
\end{equation}
The resulting imbalance
gives rise to a net CM force,
\begin{equation}
  \vec{F}_{ab} + \vec{F}_{ba}
  =
  -F_\text{nr} \, \hat{n},
  \label{eq:nonreciprocal}
\end{equation}
whose magnitude,
$F_\text{nr}
  =
  \tfrac{34}{45} \, F_0 \, k^2 (b^2 - a^2)$ ,
depends on the beads' size mismatch.
This nonreciprocal force can drive the system out of mechanical equilibrium.
It does not account for the spheres' propensity to
spin, however, because it acts along $\hat{n}$
and therefore exerts no torque. 

For simplicity, we assume that each bead
independently experiences Stokes drag with a coefficient, $\gamma_a = 6 \pi \eta_m \, a$,
that is proportional to the sphere's radius and the medium's viscosity, $\eta_m$.
Adopting the Oseen superposition
approximation yields three composite drag coefficients
for the pair of spheres in contact \cite{happel2012low,jeffrey1984calculation,felderhof2022optimizing}:
\begin{subequations}
\begin{align}
  \Gamma
  & =
    \gamma_a + \gamma_b = 6\pi\eta_m (a+b), \label{eq:Gamma} \\
  \Gamma_I
  & =
    \gamma_a \ell_a^2 + \gamma_b \ell_b^2
    =
    \Gamma \, \frac{ab \, (a+b) (a^5+b^5)}{(a^3+b^3)^2},
    \label{eq:GammaI} \\
  \Xi
  & =
    \gamma_a \ell_a - \gamma_b \ell_b
    =
    \Gamma \, \frac{ab \, (b^2 - a^2)}{a^3 + b^3} .
    \label{eq:Xi}
\end{align}
\label{eq:drag_coefficients}%
\end{subequations}
The first describes
the net CM drag.
The second characterizes rotational
drag about the CM.
The third coefficient, $\Xi$, couples
translations and rotations
\cite{happel2012low,jeffrey1984calculation,felderhof2022optimizing},
and turns out to be the origin
of intermittent spinning.

The pair's position and orientation
evolve in time under the combined influence of acoustic and
viscous forces:
\begin{subequations}
\begin{align}
  M \ddot{\vec{r}}
  & =
    -\kappa\vec{r}
    - \Gamma \, \dot{\vec{r}}
    + \Xi \, \dot{\psi} \, \hat{n}_\perp
    - F_\text{nr} \, \hat{n},
    \label{eq:eom_r} \\
  I \ddot{\psi}
  & =
    - \Gamma_I \, \dot{\psi}
    + \Xi \, \hat{n}_\perp \cdot \dot{\vec{r}}.
    \label{eq:eom_psi}
\end{align}
\label{eq:eom}%
\end{subequations}
The term proportional to $\Xi$ in Eq.~\eqref{eq:eom_psi} tends to align the
pair along its direction of travel,
as happens whenever a rigid body's center of mass and center of friction
do not coincide \cite{baconnier2025selfaligning}.
Analogous self-alignment torques describe
active polar disks \cite{weber2013long} and confined active walkers
\cite{lam2015self,dauchot2019dynamics,fersula2024self,baconnier2025selfaligning}.
Models of those self-propelled bodies have not needed the corresponding
term in Eq.~\eqref{eq:eom_r},
which couples rotations back into translations,
as the time-reversal symmetry of the Stokes mobility tensor requires.
Here it is essential, because it closes a loop that powers the pair's
activity.

The system reaches mechanical equilibrium
when nonreciprocal pair interaction
is balanced by the trap:
\begin{equation}
  F_\text{nr} \, \hat{n}
  =
  - \kappa \, r_c \, \hat{r}.
  \label{eq:equilibrium}
\end{equation}
Referring to Eq.~\eqref{eq:stiffness} and Eq.~\eqref{eq:nonreciprocal},
the equilibrium CM displacement, $r_c(a, b)$, depends on
$ka$, $kb$ and $f$, but, notably,
not on $p_0$ or properties of the medium.
The scatter plot in Fig.~\ref{fig:experiment}(a) compares measurements
of $r_c(a, b)$ for pairs of millimeter-scale
EPS beads that reach static equilibrium
with the
theoretical displacement, $r_t(a, b)$,
in an isotropic trap with $f = 1$
(Sec.~\ref{app:measuring_equilibrium}).
The observed linear trend yields
an estimate for the aspect ratio,
$f = \num{0.075(6)}$,
that agrees with orthogonal measurements
for this class of levitators \cite{marrara2024optical},
validating Eq.~\eqref{eq:nonreciprocal} for the passive state.

Linearizing Eq.~\eqref{eq:eom} about the static equilibrium
reveals that the passive state is stable if and only if the
transverse dynamics satisfy a Routh-Hurwitz
condition
(Sec.~\ref{app:static_stability}).
In the underdamped limit,
the static equilibrium is stable for pairs
of beads that satisfy
\begin{subequations}
\label{eq:stability_condition}
\begin{equation}
  S(a,b) < 1,
\end{equation}
given their dimensionless activity number,
\begin{equation}
  S(a, b)
  \equiv
  \frac{17}{45} \frac{k^4}{f} \, ab \, \Phi(ka+kb) \, (b - a)^2 .
  \label{eq:S_def}
\end{equation}
\end{subequations}

A static pair of beads neither
dissipates energy nor consumes it;
the equilibrium described by Eq.~\eqref{eq:equilibrium} is a genuinely passive state.

The system transitions to an active state through a
non-Hermitian, nonreciprocity-driven Hopf bifurcation
\cite{fruchart2021nonreciprocal,ashida2020nonhermitian} when the
Routh-Hurwitz criterion becomes an equality.
Generalizing Eq.~\eqref{eq:stability_condition} to
account for finite drag
(Sec.~\ref{app:static_stability}) shows that the
onset of instability requires $F_\text{nr}$ to exceed a
critical value,
\begin{equation}
  F_\text{nr} > F_\text{nr}^c \equiv
  \frac{\Gamma}{\Xi}
  I f \omega_0^2
  +
  \frac{\Gamma\Gamma_I - \Xi^2}{\Xi}
  \left(\frac{\Gamma}{M} + \frac{\Gamma_I}{I}\right) .
  \label{eq:Fnr_crit}
\end{equation}
This threshold diverges for spheres
of equal size, signifying that matched pairs are always passive.
More generally, it diverges whenever $\Xi$ vanishes:
$F_\text{nr}$ enters the stability condition only through the loop gain,
$F_\text{nr} \Xi$
(Sec.~\ref{app:static_stability}).

The dimensionless distance from threshold,
\begin{equation}
  \varepsilon \equiv \frac{F_\text{nr}}{F_\text{nr}^c} - 1,
  \label{eq:epsilon_def}
\end{equation}
is the natural expansion parameter for the Hopf normal form.
The complementary influence of drag is conveniently
quantified by the dimensionless damping ratio,
\begin{equation}
  \zeta \equiv \frac{\eta_m k^2}{\rho_p \, \omega_0}.
  \label{eq:zeta_def}
\end{equation}
Together, $\varepsilon$ and $\zeta$ determine the system's asymptotic dynamical state.

Just above threshold, $\varepsilon \gtrsim 0$,
the system enters a Hopf orbit (Sec.~\ref{app:hopf})
that consists of small-amplitude librations of
the absolute orientation, $\psi$,
at frequency
\begin{equation}
  \omega
  =
  \omega_0 \, \sqrt{f + \left(\frac{9}{2} \frac{\zeta}{k^2ab}\right)^2}.
  \label{eq:hopf_frequency}
\end{equation}
This orbit spontaneously breaks the continuous time-translation
symmetry of the static equilibrium in favor of the
limit cycle's discrete periodicity.
A typical observation of this rocking mode
is plotted in Fig.~\ref{fig:experiment}(b).
The amplitude of the librations
and the CM displacement both grow as
$\varepsilon$ increases.
The nonreciprocal force, $F_\text{nr}\hat{n}$, acts as a follower force,
of the kind responsible for flutter in Beck's column
\cite{beck1952} and Ziegler's pendulum \cite{ziegler1952}.
Those follower forces destabilize their systems even without damping.
This one cannot destabilize the pair on its own, because it exerts no
torque; only the reciprocal coupling, $\Xi$, returns one.
The instability is therefore enabled by dissipation.
Whereas fluttering
is a passive response to external
driving \cite{belmonte1998flutter,field1997chaotic},
these librations are an
actively sustained emergent dynamical state.

Equation~\eqref{eq:eom} also
admits orbiting solutions in which the CM
traces a circle at fixed radius, $r = r_\text{lc} > r_c$, with constant
angular velocity, $\dot\theta = \Omega$, and constant inclination,
$\varphi = \varphi_\text{lc}$.
Locking $\psi$ to $\theta$ in this way implies that the
pair pinwheels as its center orbits, realizing
the kind of spinning captured in Fig.~\ref{fig:schematic}(a),
and additionally breaking the system's discrete parity symmetry
by selecting one sense of rotation over the other.
An example is
plotted in Fig.~\ref{fig:experiment}(c).
In the weak-damping limit (Sec.~\ref{app:limit_cycles}), the spinning rate
reduces to
\begin{equation}
  \Omega
  \approx
  \omega_0
  \sqrt{f \left[ 1 +
  \frac{a^2 b^2 (a + b)}{a^5 + b^5} \,
    S(a, b) \right]},
  \label{eq:Omega_explicit}
\end{equation}
independent of the medium's viscosity.
Unlike the spinning state of confined active walkers
\cite{dauchot2019dynamics},
neither this solution nor the librational state is a stable
limit cycle.

\begin{figure}
  \centering
  \includegraphics[width=\columnwidth]{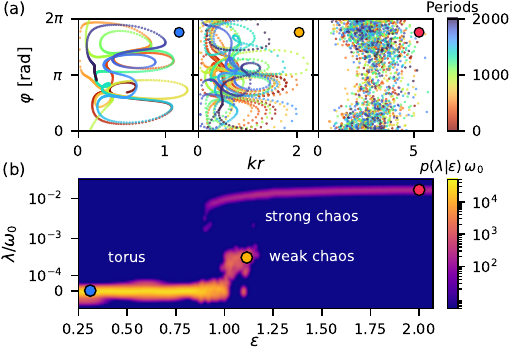}
  \caption{(a) Poincar\'e sections in the
    $(kr, \varphi)$ plane at
    $\varepsilon = \num{0.31}$ (torus), $\num{1.11}$ (weak chaos) and
    $\num{2.00}$ (strong chaos), given
    $ka = \num{0.35}$,
    $kb = \num{1.5}$ and
    $\zeta = \num{0.002}$. Points are
    colored by period number.
    (b) Probability density, $p(\lambda\,|\,\varepsilon)$, of the largest
    Lyapunov exponent, $\lambda$, (on an arcsinh-scaled axis with linear
    width \num{e-4})
    for random initial conditions.
    Plot symbols correspond to trajectories in (a).}
  \label{fig:chaosonset}
\end{figure}

Supercriticality of the Hopf bifurcation
guarantees that the librational orbit exists, with amplitude
growing continuously from zero as $\varepsilon$ increases.
A spinning orbit also exists, with amplitude that
is independent of $\varepsilon$ in the weak-damping limit
(Sec.~\ref{app:limit_cycles}).
Neither guarantee extends to the stability of these orbits.
The system's continuous rotational symmetry replicates each
orbit into a circle of equivalent orbits, which means that the system has a marginal (Goldstone)
Floquet multiplier of unit modulus.
The orbit's stability therefore depends on the remaining multipliers
\cite{golubitsky1988singularities}.
In fact, the Neimark-Sacker instability \cite{kuznetsov2004,golubitsky1988singularities}
applies to both active orbits.
Computing the monodromy matrix (Sec.~\ref{app:limit_cycles}) yields conjugate pairs of
Floquet multipliers, $\mu_j$, whose modulus exceeds
unity, identifying both orbits as saddles
rather than attractors.
Each orbit's
frequency and multiplier
correspond to a characteristic lifetime \cite{kuznetsov2004},
\begin{equation}
  \tau_j = \frac{2\pi}{\omega_j \, \ln\abs{\mu_j}},
  \label{eq:lifetime}
\end{equation}
where $(\omega_j, \mu_j) = (\omega, \mu_\text{libration})$
for librations and
$(\Omega, \mu_\text{NS})$ for spinning.
Because $\tau_j$ is an $e$-folding time, a perturbation of
relative size $\delta$ takes $\tau_j \ln(1/\delta)$ to grow to
order unity.
For the reference pair at the experimental drive and damping,
$\tau_\text{libration} \approx \qty{1.7}{\second}$ and
$\tau_\text{spinning} \approx \qty{0.1}{\second}$ at $\varepsilon = \num{0.5}$,
and simulated reversals on the torus are separated by about
\qty{4}{\second}, comparable to the intermittent episodes
plotted in Fig.~\ref{fig:schematic}(b).

Trajectories that depart from
either orbit spiral outward and settle onto a stable invariant
two-torus in phase space that combines the orbital frequency with
a slower modulation.
Physically, the CM orbits the
trap center while the pair's orientation, $\psi$, rocks back and forth
synchronously.
Episodes of spinning are transits of the torus past the unstable
spinning orbit.
The episodic nature of active transits on the quasiperiodic torus
is evident in the first Poincar\'e section
plotted in Fig.~\ref{fig:chaosonset}(a)
(Sec.~\ref{app:numerics}).

Numerical measurement of Lyapunov exponents establishes that the
quasiperiodic torus is an asymptotic dynamical state
of the system for $0 \lesssim \varepsilon \lesssim 1$.
The same method establishes that chaos is another asymptotic destination
when the system is further from equilibrium, consistent with the
Ruelle-Takens-Newhouse route to chaos via quasiperiodicity
\cite{ruelle1971nature,newhouse1978occurrence}.
Why one state gives way to the other as $\varepsilon$ increases
follows from the same instability that creates the torus in the
first place: the modulus of the librational orbit's unstable Floquet
multiplier, $\abs{\mu_\text{libration}}$, grows steadily with
$\varepsilon$, driving trajectories through
increasingly large, increasingly nonlinear excursions that a
quasiperiodic torus cannot indefinitely survive.
Where the breakdown occurs is determined numerically
from the largest Lyapunov exponent, $\lambda$
(Sec.~\ref{app:numerics}).

The last Poincar\'e section in
Fig.~\ref{fig:chaosonset}(a) shows
no sign of regular orbits.
Figure~\ref{fig:chaosonset}(b) quantifies this transformation
with the probability density, $p(\lambda\,|\,\varepsilon)\,\omega_0$, of the largest
Lyapunov exponent, $\lambda$, for random initial conditions.
Below $\varepsilon \approx \num{0.9}$, every initial condition
relaxes onto the torus, for which $\lambda$ vanishes to within
our resolution.
Strong chaos appears abruptly near $\varepsilon \approx \num{0.9}$,
with $\lambda \approx \num{5e-3} \, \omega_0$.
Initial conditions that would otherwise remain on the torus
instead find weakly chaotic states near $\varepsilon \approx \num{1.0}$,
with $\lambda \approx \num{3e-4} \, \omega_0$,
that nevertheless remain organized.
Strong chaos captures a growing fraction of initial conditions
until it is the only outcome for $\varepsilon \gtrsim \num{1.2}$.
Near its onset, the pair keeps its sense of rotation indefinitely;
farther from equilibrium, reversals recur at unpredictable times,
despite the absence of external noise sources.
The largest Lyapunov exponent rises smoothly with $\varepsilon$.
At the $\varepsilon = \num{2.0}$
reference point, $\lambda$ converges robustly to a
non-decaying plateau, $\lambda \approx \num{1.7e-2} \, \omega_0$,
which corresponds to an
e-folding time of order \qty{0.1}{\second}.

Motivated by observations of active symmetry breaking in
acoustically levitated beads, we have introduced a minimal,
first-principles route from quiescence to deterministic chaos
based on emergent activity and reciprocal dissipation.
Chaos from nonreciprocal coupling previously has been inferred
statistically in large disordered systems of passive units, such as
asymmetric neural networks \cite{sompolinsky1988chaos,kadmon2015transition}
and random ecological communities
\cite{bunin2017ecological,galla2018dynamically},
and has been built into networks of inherently active units
\cite{bick2018chaos,fuente1999diversity,brock1998heterogeneous}.
The present study identifies a feedback loop that
allows chaos to emerge in systems with few degrees of freedom:
a nonreciprocal internal force that drives one of the system's
degree of freedom and reciprocal dissipation that closes the loop.
Activity sets in when the loop's gain exceeds its losses, and
a continuous symmetry carries it on to chaos in systems
with at least three degrees of freedom.
Nothing in this loop is specific to acoustics.
Any confined body pushed by a force that follows its orientation without
exerting a torque, with its center of drag offset from its center of
mass \cite{brenner1964stokes}, contains it,
as may levitated nanoparticles \cite{reisenbauer2024non},
robotic metamaterials \cite{brandenbourger2019non},
and, where reciprocal dissipation can be identified,
genetic circuits \cite{elowitz2000synthetic}
and social networks \cite{fruchart2021nonreciprocal}.
Because such systems supply their own noise, their
fluctuations and response can be measured directly,
and ensembles of such units
\cite{brown2026liquid} should reveal whether that
intrinsic noise organizes
collective behavior, as it does in coupled chaotic maps
\cite{miller1993macroscopic}.

\begin{acknowledgments}
  This work was supported by the National Science Foundation under
  Award No.~DMR-2428983. Gordon Ni contributed to the experimental
  measurements.
\end{acknowledgments}


\clearpage
\renewcommand{\thesection}{S\arabic{section}}
\renewcommand{\theequation}{S\arabic{equation}}
\renewcommand{\thefigure}{S\arabic{figure}}
\renewcommand{\thetable}{S\arabic{table}}
\setcounter{section}{0}
\setcounter{equation}{0}
\setcounter{figure}{0}
\setcounter{table}{0}

\begin{center}
  \textbf{\large Supplemental Material}
\end{center}

\section{Materials and Methods}
\label{app:parameters}

The measured \cite{morrell2023acoustodynamic}
mass density of the type II EPS beads
(Yiwu City Hongda Foam Co., Ltd.) used for this study
is $\rho_p = \qty{30.5}{\kg\per\cubic\meter}$.

The measured \cite{morrell2023acoustodynamic}
pressure amplitude in our implementation of the
TinyLev2 design \cite{marzo2017tinylev}
is $p_0 = \qty{1600(100)}{\pascal}$
when driven with \qty{\pm 12}{\volt},
which corresponds to an energy density, $p_0^2/(\rho_m c_m^2) = \qty{18(2)}{\joule\per\cubic\meter}$,
given $\rho_m = \qty{1.225}{\kg\per\cubic\meter}$ and $c_m = \qty{343}{\meter\per\second}$
for air.

Using these inputs, the characteristic frequency,
$\omega_0 / 2\pi \approx \qty{75}{\hertz}$
[Eq.~\eqref{eq:omega0}], and the beads are expected to oscillate in the trap at
$\sqrt{f} \, \omega_0 / 2\pi \approx \qtyrange{18}{34}{\hertz}$
for $f \in \numrange{0.06}{0.2}$.

The levitator differs from a standard TinyLev2 because it
has holes drilled in its 3D-printed end caps along the central
axis to provide optical access.
Whereas the video in Fig.~\ref{fig:schematic}(a) is recorded
from the side at an oblique angle, the data plotted in Fig.~\ref{fig:schematic}(b) and Fig.~\ref{fig:experiment} were
recorded along the axis.
Images of the beads are captured in the
$(x, y)$ plane with a video camera (Teledyne Flir, Blackfly S USB3.0) outfitted with a \qty{50}{\mm} lens, which yields
a system magnification
of \qty{13.68(3)}{\um/pixel}.
The beads' radii and positions in each frame
are measured to within \qty{0.4}{pixel}
using a custom-trained variant of the
YOLOv11n machine-learning object-detection framework \cite{khanam2024yolov11}.
The location of the trap is calibrated to
within \qty{0.1}{pixel} by measuring
the equilibrium position of \num{12} beads of different sizes levitated in the trap.
Using this reference, each pair's CM displacement
from the trap center, $\vec{r}_c$, is measured with a precision of \qty{0.2}{pixel}, averaged over several video frames.

\section{Equations of motion}
\label{app:eom}

In the coordinate system defined in Fig.~\ref{fig:schematic}(c),
the beads' center of mass is located at
distance, $r$, from the center of the acoustic
trap and at inclination angle, $\theta$,
relative to the (arbitrary) $\hat{x}$ axis.
The beads' centers are directed along $\hat{n}$,
which is inclined at angle $\varphi$ relative
to $\hat{r}$.
In this coordinate system,
the beads' centers are located at $\vec{r}_a = \vec{r} - \ell_a\hat{n}$ and
$\vec{r}_b = \vec{r} + \ell_b\hat{n}$,
where
\begin{subequations}
\begin{align}
  \ell_a & = \frac{b^3(a+b)}{a^3+b^3} \label{eq:ellA} \quad \text{and} \\
  \ell_b & = \frac{a^3}{b^3} \ell_a
  \label{eq:ellB}
\end{align}
\label{eq:levers}%
\end{subequations}
are the distances of the beads'
centers from the CM.
The pair's moment of inertia about its CM is then
\begin{equation}
  I = M \ell_a \ell_b.
  \label{eq:moi}
\end{equation}

In terms of the coordinates $(r, \theta, \varphi)$,
with $\varphi = \psi - \theta$,
Eq.~\eqref{eq:eom} expands into three coupled equations of motion:
\begin{subequations}
\begin{align}
  M(\ddot{r} - r\dot\theta^2)
  & =
    -\kappa r - F_\text{nr} \, \cos\varphi
    - \Gamma \, \dot{r}
    - \Xi \, \dot\psi\sin\varphi,
    \label{eq:eom_radial} \\
  M(r \ddot{\theta} + 2 \dot{r} \dot{\theta})
  & =
    -F_\text{nr} \, \sin\varphi
    - \Gamma r \, \dot{\theta}
    + \Xi \, \dot{\psi} \, \cos\varphi,
    \label{eq:eom_theta} \\
  I \ddot{\psi}
  & =
    \Xi \, (r \dot{\theta} \, \cos\varphi - \dot{r} \, \sin\varphi)
    - \Gamma_I \, \dot{\psi}.
    \label{eq:eom_phi}
\end{align}
\label{eq:eom_polar}%
\end{subequations}
Numerical simulations of the trajectories are
performed with these equations of motion.

\section{Measuring the equilibrium fixed point}
\label{app:measuring_equilibrium}

This condition requires the CM to be
displaced by
\begin{subequations}
\begin{equation}
  r_c(a, b)
  =
  \frac{1}{f} \, r_t(a, b),
\end{equation}
where
\begin{multline}
  r_t(a, b)
  =
  \frac{17}{45}
  \frac{k^4 a^3 b^3}{a^2 - ab + b^2}
  \Phi(ka + kb)
  \, (b - a)
  \label{eq:r_t}
\end{multline}
  \label{eq:r_c}%
\end{subequations}
is the limiting
displacement that would be observed in an isotropic trap.
Equation~\eqref{eq:equilibrium} predicts
that the CM will be displaced along $-\hat{n}$, with the small bead outward.

\section{Linear stability of the equilibrium fixed point}
\label{app:static_stability}

To assess the stability of the static configuration,
we linearize Eq.~\eqref{eq:eom} around
the fixed point at
$\vec{r} = (-r_c, 0)$
and $\psi = 0$.
Perturbations along $\hat{n}$ decouple from motions
in other directions and give rise
to damped harmonic oscillations.
Perturbations transverse to $\hat{n}$
are described by the state vector
$\vec{u} =
(\delta y, \dot{\delta y}, \delta\psi, \dot{\delta\psi})^\top$ and evolve
according to $\dot{\vec{u}} = \tensor{A}\vec{u}$, where
\begin{equation}
  \tensor{A} = \begin{pmatrix}
    0           & 1           & 0                    & 0           \\
    -\kappa/M   & -\Gamma/M   & -F_\text{nr}/M   & \Xi/M       \\
    0           & 0           & 0                    & 1           \\
    0           & \Xi/I       & 0                    & -\Gamma_I/I
  \end{pmatrix}.
  \label{eq:Amat}
\end{equation}
One off-diagonal component,
$A_{2,3} = -F_\text{nr}/M$, couples the pair's tilt, $\delta\psi$, to the transverse component of the
CM force.
This is a \emph{follower force}
of the kind that is responsible for flutter instabilities in Beck's
column \cite{beck1952} and Ziegler's pendulum \cite{ziegler1952},
among other dynamical systems \cite{bolotin1963nonconservative,kirillov2021nonconservative}.
This term
enters $\tensor{A}$ asymmetrically ($A_{2,3} \neq A_{3,2} = 0$), which is the hallmark
of circulatory contribution to the system's dynamics. 
The remaining off-diagonal pair, $A_{2,4} = \Xi/M$ and $A_{4,2} = \Xi/I$, comprises the
\emph{coupling tensor}
that Brenner demonstrated can impart a steady
spinning motion to clusters of
sedimenting particles \cite{brenner1964stokes}.
The static state's stability is dictated by the signs of the
real parts of the eigenvalues of $\tensor{A}$.

One eigenvalue of $\tensor{A}$ vanishes identically because
$\det\tensor{A} = 0$.
The associated eigenvector,
$(-r_c \delta\psi, \, 0, \, \delta\psi, \, 0)^\top$,
corresponds to a rigid rotation of the static equilibrium that
costs no energy
and thus is a
Goldstone mode of the system's continuous rotational symmetry.
This soft mode has $\dot{\delta\psi} = 0$ and therefore
does not describe spinning.

The remaining three
eigenvalues are roots of the cubic equation,
\begin{equation}
  s^3 + \alpha\, s^2 + \beta\, s + \gamma = 0,
  \label{eq:cubic}
\end{equation}
with coefficients
\begin{subequations}
\begin{align}
  \alpha
  & =
    \frac{\Gamma}{M} + \frac{\Gamma_I}{I}, \label{eq:alpha}\\
  \beta
  & =
    \frac{\gamma_a \gamma_b (a+b)^2}{MI} + \frac{\kappa}{M},
    \label{eq:beta}\\
  \gamma
  & =
    \frac{F_\text{nr} \, \Xi + \kappa \, \Gamma_I}{MI}.
    \label{eq:cgamma}
\end{align}
\label{eq:rh_coeffs}%
\end{subequations}
Equation~\eqref{eq:beta} follows from the identity
\begin{equation}
  \Gamma\Gamma_I - \Xi^2 = \gamma_a \gamma_b (a+b)^2,
  \label{eq:identity}
\end{equation}
which in turn follows from expanding
$\Gamma\Gamma_I - \Xi^2 = \gamma_a \gamma_b (\ell_a + \ell_b)^2$
and noting that $\ell_a + \ell_b = a + b$.
By the Routh-Hurwitz criterion~\cite{strogatz2015},
the non-rotating equilibrium is
linearly stable if and only if $\alpha > 0$, $\gamma > 0$, and
\begin{equation}
  \alpha\beta > \gamma.
  \label{eq:rh}
\end{equation}
The first two conditions hold for all $b \geq a > 0$.
Equation~\eqref{eq:rh} therefore is
the sole stability criterion
for the equilibrium state.

Written as second-order equations for $(\delta y, \delta\psi)$,
the stiffness couplings are one-way:
tilting the pair produces a transverse force, $-F_\text{nr}\,\delta\psi$,
but no displacement produces a torque.
The follower force therefore cannot destabilize the equilibrium by itself:
for $\Xi = 0$ the tilt decouples and relaxes, and the remaining
eigenvalues have negative real parts.
Instead, $F_\text{nr}$ enters Eq.~\eqref{eq:cubic} only through $\gamma$,
in the product $F_\text{nr}\Xi$, which is the gain of the loop that
carries a tilt into a transverse force and, through drag, back into a
torque.
A negative loop, $F_\text{nr}\Xi < 0$, would destabilize the equilibrium
through a real eigenvalue rather than a Hopf bifurcation;
here both factors are proportional to $b^2 - a^2$, so the loop is
positive.

\begin{figure}
  \centering
  \includegraphics[width=0.85\columnwidth]{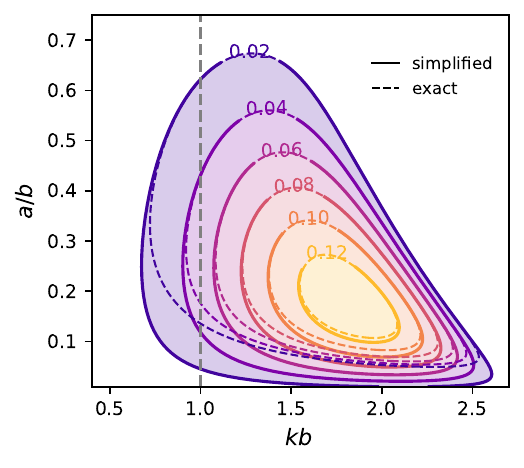}
  \caption{Stability conditions for emergent pair spinning.
    Solid curves and shaded regions show the sufficient, underdamped
    condition of Eq.~\eqref{eq:stability_criterion}; dashed curves show the exact Routh-Hurwitz condition,
    Eq.~\eqref{eq:rh}, evaluated at the bare-Stokes damping ratio,
    $\zeta = \num{6.7e-4}$, for the medium and particle
    properties used throughout this paper.
    Both the exact and limiting stability domains
    are contoured over
    the values of $f$ specified in the labels.
    Static equilibrium is expected for beads whose
    radii fall outside the relevant contour.
    The vertical dashed line at $k b = 1$ bounds
    the region of validity of the Rayleigh approximation.}
  \label{fig:stability}
\end{figure}

The most stringent form of the stability condition is obtained
by taking the limit of vanishing viscosity.
In this limit,
the static equilibrium is stable for particles
whose radii satisfy
\begin{subequations}
\begin{equation}
  S(a, b) < 1,
  \label{eq:onset_geometric}
\end{equation}
where the pair's activity number is
\begin{equation}
  S(a, b)
  \equiv
  \frac{17}{45} \frac{k^4}{f} \, ab \, \Phi(ka + kb) \, (b - a)^2 .
\end{equation}
\label{eq:stability_criterion}%
\end{subequations}
Both the exact and approximate stability
conditions are plotted in
Fig.~\ref{fig:stability}.

The geometric condition in Eq.~\eqref{eq:stability_criterion} is
sufficient to ensure that the static state is stable.
The complementary condition, however, does not
guarantee that a pair will spin.
Pairs with $S > 1$ may still reach static equilibria
that are stabilized by drag in media
with sufficiently high
viscosity.
The exact Routh-Hurwitz condition, Eq.~\eqref{eq:rh}, therefore
remains the definitive test of whether a given pair
can spin in a given
medium.

At the Hopf bifurcation, the cubic characteristic equation
Eq.~\eqref{eq:cubic}, factors as $(s^2 + \beta)(s + \alpha)$, yielding a
pair of purely imaginary eigenvalues, $s = \pm i \, \omega$, where
\begin{equation}
  \omega
  =
  \sqrt{\beta}
  =
  \sqrt{\frac{\gamma_a \gamma_b (a+b)^2}{MI} + f \omega_0^2},
  \label{eq:omega_Hopf}
\end{equation}
sets the characteristic frequency scale for emergent
dynamics.
Although the underlying instability is powered by the
particles' nonreciprocal interactions,
$\omega$ depends only on passive properties
of the system, most notably the drag experienced by the
moving beads at onset.
Drag increases the Hopf frequency relative to the trap's
natural frequency by coupling two independently damped
degrees of freedom into the system's transverse dynamics,
a recognized effect in nonconservative circulatory systems
\cite{herrmann1965destabilizing,kirillov2005stabilization,krechetnikov2007dissipation}.

\section{Supercriticality of the Hopf bifurcation}
\label{app:hopf}

We compute the first Lyapunov coefficient $\lambda_1$ via center
manifold reduction~\cite{kuznetsov2004}: $\lambda_1 < 0$ implies
supercriticality.
The system's Hopf dynamics unfold in the transverse plane spanned by
$\vec{u} = (\delta y, \dot{\delta y}, \delta\psi, \dot{\delta\psi})^\top$
and described by the system of equations
\begin{subequations}
\begin{align}
  M \ddot{\delta y}
  & =
    -\kappa \, \delta y
    - F_\text{nr} \sin(\delta\psi)
    - \Gamma \, \dot{\delta y}
    + \Xi \, \dot{\delta\psi} \cos(\delta\psi),
  \label{eq:transverse_y}\\
  I \ddot{\delta\psi}
  & =
    \Xi \, \dot{\delta y} \cos(\delta\psi)
    - \Gamma_I \, \dot{\delta\psi}.
  \label{eq:transverse_psi}
\end{align}
\label{eq:transverse}%
\end{subequations}
Expanding $\sin(\delta\psi) = \delta\psi - (\delta\psi)^3/6 + \cdots$
and $\cos(\delta\psi) = 1 - (\delta\psi)^2/2 + \cdots$,
the linear terms reproduce $\tensor{A}\vec{u}$, and the leading
nonlinear corrections,
\begin{subequations}
\begin{align}
  M \ddot{\delta y}\big|_{\text{nl}}
  & =
    \tfrac{1}{6} F_\text{nr}\, (\delta\psi)^3
    - \tfrac{1}{2} \Xi \, \dot{\delta\psi} \, (\delta\psi)^2
    + \cdots,
  \label{eq:nl_y}\\
  I \ddot{\delta\psi}\big|_{\text{nl}}
  & =
    -\tfrac{1}{2} \Xi \, \dot{\delta y} \, (\delta\psi)^2
    + \cdots ,
  \label{eq:nl_psi}
\end{align}
\label{eq:nl}%
\end{subequations}
are both cubic.
The transverse subsystem therefore contains no quadratic nonlinearities.
In Kuznetsov's notation, this means that the bilinear form,
$\tensor{B}$, vanishes,
and the first Lyapunov coefficient simplifies to
\cite{kuznetsov2004}
\begin{equation}
  \lambda_1
  =
  \frac{1}{2\omega}
  \real{
    \vec{s}^{\top} \vec{C}(\vec{q}, \vec{q}, \vec{q}^*)
  } ,
  \label{eq:l1}
\end{equation}
where $\vec{q}$ is the right eigenvector of $\tensor{A}$
with eigenvalue $+i\omega$, $\vec{q}^*$ its complex conjugate
with eigenvalue $-i\omega$,
$\vec{s}$ is the left eigenvector normalized by the
bilinear condition $\vec{s}^{\,\top}\vec{q} = 1$,
and $\vec{C}(\vec{u},\vec{v},\vec{w})$
is the symmetric trilinear form of the cubic nonlinearities.
Indexing the state-vector components $j = 1,\ldots,4$ for
$(\delta y, \dot{\delta y}, \delta\psi, \dot{\delta\psi})$, the
nonzero entries of $\vec{C}$ are
\begin{subequations}
\begin{align}
  C_2(\vec{u}, \vec{v}, \vec{w})
  & =
  - \frac{\Xi}{M}
    (u_3 v_3 w_4 + u_3 v_4 w_3 + u_4 v_3 w_3)
    \nonumber \\
  & \quad +
    \frac{F_\text{nr}}{M} \, u_3 v_3 w_3
  \label{eq:C2}\\
  C_4(\vec{u}, \vec{v}, \vec{w})
  & =
  -\frac{\Xi}{I}
    (u_2 v_3 w_3 + u_3 v_2 w_3 + u_3 v_3 w_2).
  \label{eq:C4}
\end{align}
\label{eq:Ctri}%
\end{subequations}
Component $C_2$ encodes the
$-\tfrac{1}{2}\dot{\delta\psi}(\delta\psi)^2$ coupling that
arises when $\cos(\delta\psi)$ is expanded (the $\Xi$ term)
and the $(\delta\psi)^3/6$ correction
to $\sin(\delta\psi)$ (the $F_\text{nr}$ term in Eq.~\eqref{eq:C2}).
Component $C_4$ encodes the analogous
$-\tfrac{1}{2}\dot{\delta y}(\delta\psi)^2$ coupling in the torque
equation.
In the underdamped limit, $\Gamma/M \ll \omega$ and $\Gamma_I/I \ll \omega$,
the contraction $\vec{s}^{\,\top}\vec{C}(\vec{q},\vec{q},\vec{q}^*)$
is proportional to $-F_\text{nr} \, \Xi^3$.
Since $F_\text{nr} > 0$ and $\Xi > 0$ for all $b > a$, we
obtain $\lambda_1 < 0$ universally, which means
that the Hopf bifurcation is
supercritical for all values of the parameters.

This derivation treats the transverse subsystem, Eq.~\eqref{eq:transverse},
as closed at cubic order. In fact, the longitudinal coordinate, $\delta x$,
is weakly driven at $O(\delta\psi^2)$ and feeds back into Eq.~\eqref{eq:eom_psi} at the same
cubic order that is retained elsewhere. Because $\delta x$ remains a stable,
externally forced oscillator rather than an independent source of
instability, this coupling is fully accounted for by the standard
center-manifold construction~\cite{kuznetsov2004}. Evaluated numerically
across the damping range surveyed in Fig.~\ref{fig:phasediagram}, the
resulting correction changes $\lambda_1$ by less than \qty{0.1}{\percent} at the
lightest damping tested and by \qty{9}{\percent} at the heaviest,
and never changes its sign.
The Hopf bifurcation is supercritical throughout.

\section{Limit-cycle derivation and stability}
\label{app:limit_cycles}

Setting $\dot{r} = \ddot{r} = \dot{\varphi} = 0$
and $\dot{\psi} = \Omega$ in
Eq.~\eqref{eq:eom} yields three algebraic conditions:
\begin{subequations}
\begin{align}
  r_\text{lc} \, \cos\varphi_\text{lc}
  & =
    \frac{\Gamma_I}{\Xi},
    \label{eq:lc_rot} \\
  F_\text{nr} \, \sin\varphi_\text{lc}
  & =
    \frac{\Omega}{r_\text{lc}}
    \left(\Gamma_I - \Gamma r_\text{lc}^2\right),
    \label{eq:lc_tang} \\
  M r_\text{lc} \Omega^2
  & =
    \kappa r_\text{lc}
    + F_\text{nr} \, \cos\varphi_\text{lc}
    + \Xi \, \Omega \, \sin\varphi_\text{lc}.
    \label{eq:lc_rad}
\end{align}
\label{eq:lc}%
\end{subequations}
Whereas the static pair is oriented along $\hat{n} = -\hat{r}$,
the orbiting pair flips to $\hat{n} \approx \hat{r}$ so their
nonreciprocal interaction can supplement the trap's
centripetal force.

The right-hand side of Eq.~\eqref{eq:lc_tang} scales as
$\order{\eta_m \Omega}$ and therefore is small
when damping is weak.
Taking the weak-damping limit reduces the pair's inclination ($\sin\varphi_\text{lc} \to 0$) so that
Eqs.~\eqref{eq:lc_rot} and \eqref{eq:lc_rad} reduce to
\begin{subequations}
\begin{align}
  r_\text{lc}
  & \approx
    \frac{\Gamma_I}{\Xi}
    =
    \frac{a^5 + b^5}{(a^3 + b^3) (b - a)} ,
    \label{eq:rlc} \\
  \Omega^2
  & \approx
    \frac{\kappa}{M} + \frac{F_\text{nr} \, \Xi}{M \Gamma_I}
    = f \omega_0^2 \left(1 + \frac{r_c}{r_\text{lc}}\right) .
  \label{eq:Omega_lc}
\end{align}
\end{subequations}
In this limit, the CM offset
of an orbiting pair
is independent of the nonreciprocal drive,
which instead contributes to the
orbital rate, $\Omega$.
Having located these limit cycles algebraically, we now examine their
stability.

The monodromy matrix of a $T$-periodic orbit $\vec{z}_0(t)$ of an autonomous
system $\dot{\vec{z}} = \vec{F}(\vec{z})$ is the state-transition matrix
$\tensor{M} \equiv \tensor{\Phi}(T)$ of the variational equation
$\dot{\tensor{\Phi}} = \tensor{J}(\vec{z}_0(t))\,\tensor{\Phi}$,
with initial condition
$\tensor{\Phi}(0) = \tensor{1}$, where
$\tensor{J} \equiv \partial \vec{F}/\partial \vec{z}$ is the Jacobian of the
equations of motion evaluated along the orbit
and $\tensor{1}$ is the \numproduct{6 x 6} identity tensor \cite{strogatz2015}.
Its eigenvalues are the Floquet
multipliers, $\mu_i$, that describe the rates at which
small perturbations decay to the orbit.
The orbit is linearly stable if and only if
$\abs{\mu_i} \leq 1$ for every multiplier other than the trivial $\mu = 1$
that every autonomous periodic orbit possesses along its own flow
direction~\cite{strogatz2015}.
Differentiating Eq.~\eqref{eq:eom} yields
this Jacobian in block form,
\begin{equation}
  \tensor{J} =
  \begin{pmatrix}
    0 & 0 & \tensor{1} & 0 \\
    0 & 0 & 0 & 1 \\
    -\dfrac{\kappa}{M} \, \tensor{1} &
    -\dfrac{1}{M}\big(F_\text{nr} \, \hat{n}_\perp
    + \Xi\dot\psi \, \hat{n}\big) &
    -\dfrac{\Gamma}{M} \, \tensor{1} &
    \dfrac{\Xi}{M} \, \hat{n}_\perp \\
    0 & -\dfrac{\Xi}{I} \big(\hat{n}\cdot\dot{\vec{r}}\big) &
                                                              \dfrac{\Xi}{I} \, \hat{n}_\perp^{\top} &
                                                                                                   -\dfrac{\Gamma_I}{I}
  \end{pmatrix},
  \label{eq:jac_general}
\end{equation}
with rows and columns ordered $(\vec{r}, \psi, \dot{\vec{r}}, \dot\psi)$,
and where $\hat{n}$ and $\hat{n}_\perp$ are evaluated
along the orbit at $\psi = \psi_0(t)$, which is
generic to any periodic orbit of
Eq.~\eqref{eq:eom}, librational or spinning alike.
When Eq.~\eqref{eq:jac_general} is evaluated at the static equilibrium
($\dot{\vec r}=0$, $\dot\psi=0$) instead of along an orbit, it reduces to
the transverse block of $\tensor{A}$ from Eq.~\eqref{eq:Amat}, as it must,
since both are the same Jacobian linearized about different solutions of
the same equations of motion.

Computing this monodromy matrix at the reference geometry used throughout
this work ($ka = \num{0.35}$, $kb = \num{1.5}$,
$\zeta = \num{0.002}$) confirms
that the librational orbit
is a saddle rather than
an attractor.
A complex-conjugate Floquet pair with modulus
$\abs{\mu_\text{libration}} > 1$ is already present at $\varepsilon =
\num{0.02}$, just above onset, and climbs smoothly and monotonically to
$\abs{\mu_\text{libration}} = \num{1.028}$
by $\varepsilon = \num{0.5}$.
This is a weak instability near onset, consistent with
librations being a small-amplitude structure born continuously at
threshold, and is confirmed by the correctness check (Sec.~\ref{app:numerics}).

The same construction applied to the spinning limit cycle shows that
its monodromy matrix has a richer spectral structure: one Goldstone
multiplier ($\mu = +1$, corresponding to phase shifts along the
orbit), one strongly contracting mode, one weakly stable
complex-conjugate pair, and one unstable complex-conjugate pair with
modulus $\abs{\mu_\text{NS}} > 1$, which is the signature of a
Neimark-Sacker (torus) instability in the Poincar\'e
map~\cite{strogatz2015}. Sweeping the size ratio $b/a$ from
\numrange{1.4}{10} and the forcing strength $f \in (\num{0.02},
\num{1}]$ for every combination past the Routh-Hurwitz threshold
confirms $\abs{\mu_\text{NS}} \in [\num{1.51}, \num{1.88}]$ at fixed
damping ratio $\zeta = \num{0.002}$.
This is a far stronger instability
than the librational orbit's, even near onset, consistent with the
finite amplitude required for the spinning solution.
Perturbations
transverse to the orbit spiral outward along the unstable
Neimark-Sacker directions, identifying the orbit as a saddle rather
than an attractor. This instability weakens as $\zeta \to 0$: $\abs{\mu_\text{NS}}$
approaches unity from above, though the limit is singular, since
$\Xi \to 0$ decouples $\psi$ from the translational dynamics
entirely at $\zeta = 0$.

\begin{figure}
  \centering
  \includegraphics[width=\columnwidth]{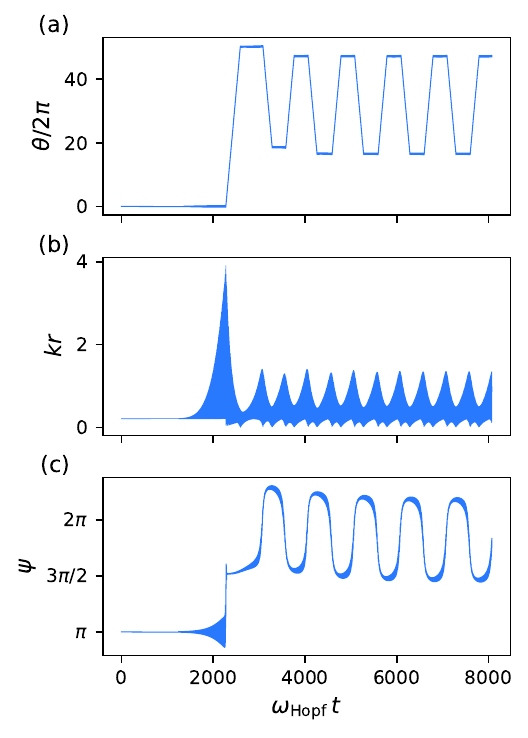}
  \caption{Typical trajectory along the quasiperiodic torus
  seeded near the unstable static equilibrium at
    ($ka = \num{0.35}$, $kb = \num{1.5}$,
    $\zeta = \num{0.002}$, $\varepsilon = \num{0.5}$).
        (a) Orbital winding number
    $\theta/2\pi$: after an initial librational stage of small-amplitude
    oscillation about equilibrium, the trajectory grows through a large
    transient overshoot and settles, by $\omega t\approx\num{2560}$, onto a
    regular sawtooth of directed rotation, brief pauses, and reversals.
    (b) The CM displacement $kr$ over the same window, peaking at each
    reversal and dipping during each pause, consistent with the
    text's description of the settled motion. (c) The pair's
    orientation, $\psi$, over the same window. Whereas the CM orbits the center of the trap in a direction that reverses at regular intervals, the pair's orientation only rocks back and forth.}
  \label{fig:torustrajectory}
\end{figure}

Trajectories that spiral away from the unstable spinning orbit settle onto
this same quasiperiodic torus: the CM winds around the trap, reversing
direction at regular intervals
[Fig.~\ref{fig:torustrajectory}(a,b)]. The pair's own orientation does
not wind up with it: $\psi$ remains confined to less than one full
turn throughout [Fig.~\ref{fig:torustrajectory}(c)], and the spin
rate $\Omega$ stays far below its value on the algebraic solution
[Eq.~\eqref{eq:Omega_explicit}], oscillating instead near zero.
The expression in Eq.~\eqref{eq:rlc} for $r_\text{lc}$
sets the
observable scale of the CM displacement's excursions, while
$\tau_\text{spinning}$
[Eq.~\eqref{eq:lifetime}]
sets the characteristic interval between
successive departures from the algebraic solution.

Distinguishing a genuine torus from marginal chaos requires long
integration. The standard renormalization
method~\cite{benettin1980lyapunov} for the largest Lyapunov exponent
$\lambda$, evaluated over a short window (of order \num{e4} time
units), returns a deceptively clean, small positive value at
essentially every point past threshold, which is indistinguishable from
chaos.
On a genuine torus, however, the running estimate
does not plateau but decays roughly as
$1/t$~\cite{benettin1976kolmogorov,goldhirsch1987stability},
and only very long
integration (\numrange{e5}{e6} time units) resolves whether it is
converging toward zero or toward a true plateau.
At $\varepsilon = \num{0.5}$, $\zeta = \num{0.002}$, the running estimate
falls from $\num{1.7e-3}\,\omega_0$ at $t = \num{1.5e4}$ to
$\num{7.9e-5}\,\omega_0$ at $t = \num{3e5}$, with no sign of
leveling off, which is the signature of a genuine torus.
Direct inspection
confirms this: the winding number and CM displacement reproduce
the intermittent, reversing orbital motion already described for the
librational and spinning orbits, but now in a completely deterministic, quasiperiodic
rhythm, while the pair's own spin rate stays small throughout and
does not participate in this slow modulation.

The seeding, renormalization, and robustness protocol behind this
estimate (Sec.~\ref{app:numerics}) is insensitive to the direction
of the initial perturbation, with five
independent choices reproducing the values quoted above at both
reference times.

The reversal period on the torus follows the same functional form as
Eq.~\eqref{eq:lifetime}, applied with the librational branch's
multiplier but using the spinning orbit's
own period as the natural timescale, even though the trajectory
departs through the librational doorway (the unstable librational orbit,
through which trajectories leave the static equilibrium):
\begin{equation}
  \tau_\text{torus} \sim c\, \frac{2\pi}{\Omega \, \ln\abs{\mu_\text{libration}}},
  \label{eq:tau_torus}
\end{equation}
with an order-unity prefactor $c \approx \numrange{2}{3}$ that is robust against
$\varepsilon$ at fixed $\zeta$. Its $\zeta$-dependence is known only
qualitatively: the period shortens as $\zeta$ increases, consistent
with the general trend that heavier damping accelerates the
instability.

\section{Activity of subcritical transients}
\label{sec:subcritical_activity}

Persistent spinning is not confined to conditions where spinning is
the asymptotic outcome.
Within the static-equilibrium region of
Fig.~\ref{fig:phasediagram} ($\varepsilon < 0$, where the
Routh-Hurwitz condition holds), a spinning trajectory
can persist for many cycles before decaying to rest,
with the number of orbits growing as
$\varepsilon$ approaches the Hopf threshold from below.
The constant-inclination orbit therefore organizes transient
spinning on both sides of onset: as a saddle that a supercritical
trajectory eventually departs, and as an unstable structure from
which a subcritical trajectory relaxes.

What powers a subcritical transient?
The spinning pair has an instantaneous mechanical energy
\begin{equation}
  E
  =
  \frac{1}{2} M \abs{\dot{\vec{r}}}^2
  + \frac{1}{2} \kappa r^2
  + \frac{1}{2} I\dot\psi^2 .
\end{equation}
Equation~\eqref{eq:eom} explains how
nonconservative forces change that energy:
\begin{equation}
  \dot{E}
  =
  -F_\text{nr} \, (\hat{n}\cdot\dot{\vec{r}})
  -
  \left[
    \Gamma\abs{\dot{\vec{r}}}^2 + \Gamma_I\dot\psi^2
    - 2\,\Xi\,\dot\psi\,(\hat{n}_\perp\cdot\dot{\vec{r}})
  \right].
  \label{eq:power_balance}
\end{equation}
The bracketed term is positive semi-definite
for any physical resistance tensor.
This follows from Eq.~\eqref{eq:identity},
which establishes that
$\Gamma\Gamma_I - \Xi^2 = \gamma_a\gamma_b(a+b)^2 \geq 0$.
Physically, this means that drag can only remove
energy from the pair.
The nonreciprocal term, by contrast, can have either sign.
Integrating it along simulated subcritical trajectories
($\varepsilon < 0$) shows that it is always positive,
which means that $F_\text{nr}$ does positive work
even under conditions where drag eventually
brings the system to rest.
In this sense, the subcritical transient is
active.

This places transient spinning in a category distinct from both the
passive equilibrium and the sustained active state on the quasiperiodic
torus: a strictly passive transient ($F_\text{nr} = 0$)
decays on its intrinsic damping timescale, with orbiting and
pinwheeling only weakly cross-coupled through $\Xi$ alone.
The transient active states probed in the
static-equilibrium region of Fig.~\ref{fig:phasediagram} are
extended in duration and have orbiting and pinwheeling locked
together by the much larger nonreciprocal coupling, yet still
eventually decay because their positive work never fully replaces
what drag removes.

\section{Multistability at the onset of chaos}
\label{app:chaos_multistability}

The boundary between the quasiperiodic torus and strong chaos is not a
single curve in $(\varepsilon,\zeta)$: because the librational and
spinning doorways are both always accessible, a trajectory that departs
through one can destabilize into strong chaos at a different $\varepsilon$
than one departing through the other. This appendix locates that
boundary for each doorway, characterizes how sharply the torus
breaks down, and gives the quantitative mechanism responsible for
the resulting multistability.

\begin{figure}
  \centering
  \includegraphics[width=\columnwidth]{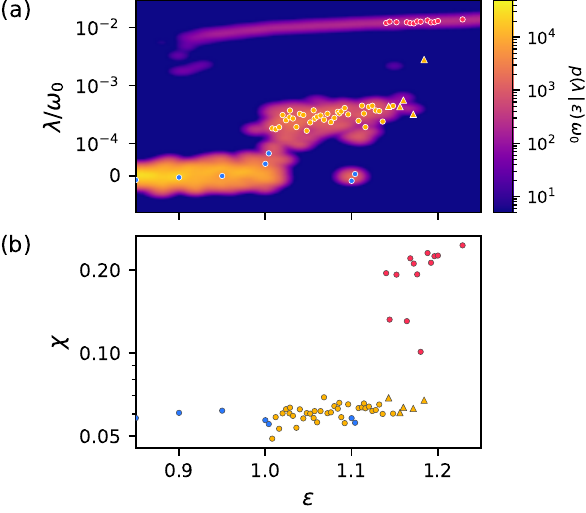}
  \caption{(a) Probability density, $p(\lambda\,|\,\varepsilon)\,\omega_0$, of the
    largest Lyapunov exponent, $\lambda$ (linear below $10^{-4}$),
    for random initial conditions, as in Fig.~\ref{fig:chaosonset}(b),
    with $\lambda$ for trajectories seeded near the static equilibrium (points),
    colored by regime: torus (blue), weak chaos (amber) and strong chaos (red).
    Triangles mark weak chaos that later escapes to strong chaos.
    (b) Coherence, $\chi$, of the same trajectories, for
    $ka = \num{0.35}$, $kb = \num{1.5}$ and $\zeta = \num{0.002}$.
    Both signal strong chaos at the same values of $\varepsilon$;
    only $\lambda$ resolves weak chaos.}
  \label{fig:chi_lambda}
\end{figure}

Mapping this boundary across $(\varepsilon, \zeta)$
(Fig.~\ref{fig:phasediagram}) with Lyapunov exponents is costly,
so we use a faster diagnostic, the coherence of a stroboscopic
Poincar\'e section sampled once per reference period,
\begin{equation}
    \chi(\varepsilon)
    =
    \frac{\operatorname{median}_i \sqrt{\Delta x_i^2 + \Delta y_i^2}}
         {\sqrt{\operatorname{Var}(x) + \operatorname{Var}(y)}} ,
    \label{eq:coherence}
\end{equation}
where $x_i = \varphi_i / 2\pi$ and
$y_i = (r_i - r_\text{min}) / (r_\text{max} - r_\text{min})$
map the section into the unit square
and $\Delta$ indicates the change in a value over one stroboscopic
period.
The coherence is small when the section traces an organized
invariant curve and large when it fills a structureless
two-dimensional region.
For trajectories seeded near the static equilibrium,
Fig.~\ref{fig:chi_lambda} shows that $\chi > \num{0.09}$ exactly where
$\lambda$ signals strong chaos, at all \num{80} values of $\varepsilon$ sampled,
so it locates strong chaos for a given seed.
It cannot, however, distinguish weak chaos from a torus.

Locating the boundary at which $\chi$ jumps
across a range of $\zeta$, cross-checked against the
Lyapunov exponent (Sec.~\ref{app:numerics}), shows
that a trajectory destabilized through the librational branch and
one destabilized through the spinning branch cross into strong chaos at
different $\varepsilon$, with the spinning pathway getting there
first over a substantial intermediate band (for example,
$\varepsilon \in \numrange{0.90}{1.23}$ at $\zeta = \num{0.002}$).
Within this band, which pathway the system happens to pass through,
not the control parameters alone, determines whether it is found on
the torus or in strong chaos.

Scanning $\chi(\varepsilon)$ densely along the
librational-doorway pathway shows that the breakdown between torus
and strong chaos is not a slow wrinkling cascade (Fig.~\ref{fig:chi_lambda}b):
$\chi$ stays near \num{0.06} through the torus and weak-chaos
range, then rises intermittently
across
$\num{1.14} \lesssim \varepsilon \lesssim \num{1.23}$, settling
consistently above \num{0.1} only at the upper end of that range,
which coincides with the librational-pathway boundary located above,
before resuming a gentler rise to $\chi(\num{2.0}) = \num{0.47}$.
The torus survives as a recognizable, if increasingly perturbed,
structure right up to a comparatively sharp collapse, rather than
gradually dissolving.

This doorway-dependent multistability has a direct, quantitative
explanation in these same trajectories. At $\varepsilon = \num{0.90}$,
trajectories departing through either doorway explore similar,
modest ranges of the CM displacement $r$ (up to $\sim\num{1.5}$),
remaining below the seeding-independent spinning radius
$k\,r_\text{lc} \approx \num{1.95}$ [Eq.~\eqref{eq:rlc}]. By
$\varepsilon = \num{0.95}$, the spinning-seeded trajectory has
already escaped past $r_\text{lc}$, deep into the nonlinear part of
the phase space, $r \in [\num{1.17},\num{5.02}]$. At the same time,
the librational-seeded trajectory remains confined to
$r \lesssim \num{1.6}$. This gap persists until
$\varepsilon \approx \num{1.2}$, where the librational-seeded range
finally widens to match the spinning-seeded one. The spinning
doorway reaches large-$r$, strongly nonlinear phase space at
substantially smaller $\varepsilon$ than the librational doorway,
which is why it destabilizes into strong chaos first.

\section{Summary of dynamical states}
\label{app:summary}

\begin{figure}
  \centering
  \includegraphics[width=\columnwidth]{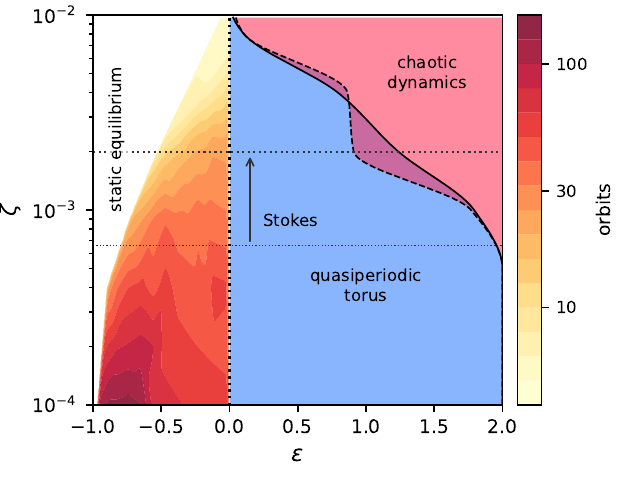}
  \caption{Dynamical states in the $(\varepsilon, \zeta)$
    plane for a specific choice of bead sizes
    ($ka = \num{0.35}$, $kb = \num{1.5}$).
    The vertical dotted line at $\varepsilon = 0$ marks
    the Hopf bifurcation between static equilibrium
    (left) and the quasiperiodic torus (blue).
    The solid and dashed curves demark transitions to strong chaos (red) via
    libration and spinning, respectively.
    The purple regions denote multistability
    where either asymptotic state remains accessible depending
    on which doorway the system passes through.
    Colors within the static-equilibrium region indicate
    the lifetime of transient spinning states.
    Horizontal dotted lines mark the Stokes and drag-corrected damping
    ratios for the TinyLev2 operating in air.}
  \label{fig:phasediagram}
\end{figure}

Figure~\ref{fig:phasediagram} presents
the state diagram in the $(\varepsilon, \zeta)$ plane for a specific pair of
levitated beads.
The Hopf bifurcation at $\varepsilon = 0$
is independent of $\zeta$ because the
transition starts from static equilibrium.
Because the bifurcation is supercritical (Sec.~\ref{app:hopf}),
the librational orbit that
emerges for $\varepsilon > 0$
grows continuously from onset;
a spinning orbit also emerges, at an amplitude
independent of $\varepsilon$ (Sec.~\ref{app:limit_cycles}).
The emergent active
state can
take two forms:
librational rocking that breaks
the continuous time-translation symmetry of the static equilibrium,
or spinning that breaks
the system's discrete parity symmetry, $\dot{\psi} \to - \dot{\psi}$.
Unlike previous studies of confined active walkers \cite{dauchot2019dynamics,baconnier2025selfaligning,damascena2022coexisting},
we find that neither mode of motion
is a stable limit
cycle.
Instead, the active state either settles asymptotically
into a quasiperiodic torus on which the orbital direction reverses
regularly,
or else descends into sustained chaos.
Table~\ref{tab:regimes} collects the states together with the
condition, key quantity, and controlling parameters for each.

\begin{table*}
  \caption{%
    Dynamical states of the acoustically levitated bead pair.
    $S(a, b)$ is the activity number
    defined in Eq.~\eqref{eq:S_def},
    $f$ is the ratio of in-plane to axial trap stiffness, and
    $\varepsilon \equiv F_\text{nr}/F_\text{nr}^c - 1$ is the reduced
    distance from the exact Routh-Hurwitz threshold.
    Only static equilibrium, the quasiperiodic torus, and chaotic
    dynamics are asymptotically stable; librations and transient
    spinning are universal, always-accessible unstable doorways
    through which the system reaches one of the three asymptotic
    states, but are not additional destinations in their own right.
    Which stable dynamical state is ultimately reached depends on
    $(\varepsilon,\zeta)$ and, within the multistable band,
    on which doorway the system passes through (Fig.~\ref{fig:phasediagram}).
  }
  \label{tab:regimes}
  \begin{ruledtabular}
    \begin{tabular}{llllll}
      State & Character & Condition & Key quantity & & Controlled by \\
      \hline
      Static equilibrium
        & asymptotically stable
        & $\varepsilon < 0$
        & $r_c$
        & Eq.~\eqref{eq:equilibrium}
        & geometry, $f$ \\
      Librations
        & unstable (doorway)
        & ---
        & $\omega$, $\tau_\text{libration}$
        & Eqs.~\eqref{eq:hopf_frequency}, \eqref{eq:lifetime}
        & geometry, $f$ \\
      Transient spinning
        & unstable (doorway)
        & ---
        & $\Omega$, $\tau_\text{spinning}$
        & Eqs.~\eqref{eq:Omega_explicit}, \eqref{eq:lifetime}
        & geometry; $\zeta$ (for $\tau_\text{spinning}$) \\
      Quasiperiodic torus
        & asymptotically stable
        & $\varepsilon$, $\zeta$ below boundary
        & $\tau_\text{torus}$
        & Eq.~\eqref{eq:tau_torus}
        & $\varepsilon$, $\zeta$, doorway \\
      Chaotic dynamics
        & asymptotically stable
        & $\varepsilon$, $\zeta$ above boundary
        & $\lambda$
        & Fig.~\ref{fig:chaosonset}
        & $\varepsilon$, $\zeta$, doorway \\
    \end{tabular}
  \end{ruledtabular}
\end{table*}

\section{Approximations}
\label{sec:approximations}

The model developed here rests on several simplifying assumptions
that limit the quantitative accuracy of its predictions and merit
comment.

\subsection{Acoustic contrast}

The expressions used
for the acoustic radiation force and K\"{o}nig interaction
\cite{king2025scattered}
are simplified by assuming the levitated particles to be
much denser and less compressible than the medium:
$\rho_p \gg \rho_m$ and $\kappa_p \ll \kappa_m$.
In this approximation, the acoustic contrast factor,
$(\rho_p - \rho_m)/(\rho_p + 2\rho_m) - \kappa_p/(3\kappa_m)$,
saturates to its maximum value of unity.
This is a reasonable approximation for EPS beads in air
because $\rho_p/\rho_m \approx \num{16}$ \cite{morrell2023acoustodynamic} and $\kappa_p/\kappa_m < \num{e-3}$.
Finite-contrast corrections nevertheless modify both
the force scale, $F_0$, and the trap stiffness, $\kappa$,
and hence the activity number, $S(a, b)$.
Accounting for smaller acoustic contrast factors would
allow spinning at smaller size asymmetries than
the present model predicts.

\subsection{Single-scattering approximation}
\label{sec:approx_scattering}

The K\"{o}nig interaction is computed in the first Born approximation:
each sphere scatters the incident field once, and the force on each
sphere is evaluated in the field produced by the unperturbed source
plus the singly-scattered field of the other sphere~\cite{king2025scattered}.
This approximation is expected to be accurate when
the spheres are well separated.
The contact geometry
that defines $F_0$ places the sphere surfaces in
close proximity, where multiple re-scattering events between the
beads may become significant.
Corrections beyond the Born approximation would modify the prefactor
and angular dependence of the contact force $F_0$.

A rough estimate of the size of this correction follows from how
quickly a Rayleigh scatterer's own re-radiated field falls off away
from its surface: at a distance $d$ from a sphere of radius $\sigma$,
the near field is suppressed, relative to the field at the sphere's
own surface, by a factor of order $(\sigma/d)^3$.
At contact, $d = a+b$, so the more consequential of the two such
factors is $[b/(a+b)]^3$, since it is the intense near field of the
\emph{larger} sphere that acts back on its smaller partner.
This factor is modest for size-matched pairs
but climbs to \num{0.5} at the size ratio
$b/a = \num{4.3}$ used throughout the numerical examples in this
work, precisely because reaching threshold at the measured aspect ratio
requires a large size disparity in the first place
(\S~\ref{sec:approx_trap}).
The comparison between theory and experiment for the
system's statics in Fig.~\ref{fig:experiment}
suggests that predictions for nonreciprocal
interparticle forces capture key qualitative features
of the size dependence, at least for modest size ratios.

The true interaction at contact is likely stronger than this leading-order,
ideal-fluid estimate. Measurements and numerical calculations of interparticle
acoustic forces at close range
\cite{mohapatra2018experimental,simon2019numerical,hoque2020interparticle,silva2026acoustic}
indicate an enhancement on the order of a factor of two, although the
underlying physics is not yet settled.

The phase diagram of Fig.~\ref{fig:phasediagram} is built
at one fixed, illustrative bead geometry that clearly exhibits
the full phenomenology:
$(ka, kb) = (\num{0.35},\num{1.5})$.
The dynamical control parameters $\varepsilon$ and $\zeta$ are
then set, in principle, by varying properties of the levitator and
the density of the bead material.
Experimental studies, by contrast, tend to vary $(ka, kb)$
while holding the other parameters constant.
This pragmatic approach offers glimpses of most of the
system's modes of motion, but does not cover nearly as much
of the dynamical phase space as the $(\varepsilon, \zeta)$ scan
presented here.
This also reduces the urgency of achieving quantitative
agreement with experiment.

\subsection{Rayleigh approximation}
\label{sec:approx_rayleigh}

The acoustic forces that contribute to the spheres'
equation of motion in Eq.~\eqref{eq:eom}
are evaluated in the Rayleigh approximation, $ka, kb < 1$ \cite{king2025scattered}.
The larger of the two beads used
for numerical studies in this work has $kb = \num{1.5}$,
which exceeds this bound, but is consistent with
the sizes of the beads in experimental observations of
intermittent spinning.
Falling back on the Rayleigh approximation greatly simplifies the
analytic expressions for acoustic forces and therefore
highlights the mechanism by which steady-state chaos can arise
in a system powered by emergent activity.

\subsection{Harmonic trap approximation}
\label{sec:approx_trap}

The confining potential, Eq.~\eqref{eq:confining_potential}, is the leading-order, parabolic
expansion of the true, periodic acoustic standing wave about a pressure node.
The linear approximation is
trustworthy for displacements smaller
than roughly a quarter of the wavelength of sound.
The reference system used for this work, however, ventures beyond
the domain of validity:
the spinning-limit-cycle radius, $k\,r_\text{lc} \approx \num{1.95}$ [Eq.~\eqref{eq:rlc}],
corresponds to \qty{2.67}{\milli\meter}, which exceeds the
quarter-wavelength bound by \qty{25}{\percent}.
In practice, nonlinearities in the confining potential are likely
to further enrich the system's phenomenology.
Sticking to the linear approximation establishes that such
effects are not required for chaos to emerge in emergently
active systems.

The reference pair lies within the experimental range of bead sizes.
At $(ka, kb) =
(\num{0.35},\num{1.46})$ [diameters of roughly \qty{1}{\milli\meter} and \qty{4}{\milli\meter}],
it is already active at the measured aspect ratio and damping,
with $\varepsilon = \num{0.38}$ and no correction to $F_0$.
Boosting $F_0$ (\S~\ref{sec:approx_scattering}) by a factor of \num{1.24}
($\varepsilon = \num{0.71}$) yields a torus whose excursions,
\qty{2.05}{\milli\meter}, remain within the Hooke's law approximation,
whereas a factor of \num{1.7} ($\varepsilon = \num{1.34}$) produces
chaos through both the librational and spinning
doorways with excursions of about \qty{8.6}{\milli\meter}, comparable to the
wavelength of sound.

This exposes a tension with the Rayleigh approximation. Satisfying the
Rayleigh bound $kb < 1$ favors nearly equal-sized beads, but the activity
number $S(a, b) \propto (kb - ka)^2$ [Eq.~\eqref{eq:S_def}] that sets the onset of
spinning [Eq.~\eqref{eq:onset_geometric}] is then small, so reaching the
instability demands a large nonreciprocal force. The equilibrium displacement
$r_c = F_\text{nr}/\kappa$ [Eq.~\eqref{eq:equilibrium}] produced by that force already
reaches the quarter-wavelength bound at the onset of spinning, so a nearly
equal-sized pair is driven out of the harmonic region of the trap before it can
spin. The large size disparity of the reference geometry is therefore not
incidental: it is what keeps the center of mass within the harmonic trap while
the pair remains active. Consistently, at the measured aspect ratio and
damping, the active pairs of reasonable size ($kb < \num{2.2}$) reach only as far
as the quasiperiodic torus:
\num{321} of \num{1404} pairs on a $(ka, kb)$ grid are active, all quasiperiodic,
with $b/a \geq \num{2.4}$ and $\varepsilon \leq \num{0.7}$.
The chaotic regime, $\varepsilon \gtrsim 1$, requires a stronger contact
interaction, by a factor of order \num{1.5}, or a smaller aspect ratio,
together with beads large enough to leave the Rayleigh regime.

The pairs in Fig.~\ref{fig:schematic}(b) and Fig.~\ref{fig:experiment}(b,c)
have $b/a$ between \num{1.04} and \num{1.6}, well below this threshold, and
the model calls all three passive. The known corrections do not close this
gap: the near-field enhancement to $F_\text{nr}$
(\S~\ref{sec:approx_scattering}, a factor of about \num{2}) leaves the two
Fig.~\ref{fig:experiment} pairs subcritical, and the rotational drag
correction (\S~\ref{sec:approx_hydro}) is of comparable size at these
near-equal geometries but raises $F_\text{nr}^c$ rather than lowering it.
What powers these pairs' observed activity is not accounted for by the
present model.

\subsection{Stokes drag}
\label{sec:approx_drag}

Viscous drag on each of the beads is modeled in the Stokes limit,
which assumes the Reynolds number to be small:
$\text{Re} = \rho_m v\sigma/\eta_m \ll 1$.
The observed spinning rates of a few hertz correspond to bead
surface speeds of order
$v \sim \Omega a \sim \qty{e-2}{\meter\per\second}$
for millimeter-scale beads, giving $\text{Re} \sim \numrange{1}{10}$.
Inertial corrections to the Stokes drag, which enter at
$\order{\text{Re}}$, are therefore not entirely negligible and
will shift the quantitative predictions of the limit-cycle amplitude
and spin rate.
Center-of-mass speeds on the torus and in chaos are higher still,
raising $\text{Re}$ into the tens; Stokes drag is at best a
qualitative approximation in these regimes.

A second, independent inertial correction arises from the
unsteadiness of the flow, distinct from the finite-Reynolds-number
effect above.
The beads' secular dynamics,
libration, spinning, and the reversals between them,
evolve on the timescale set by the trap
frequency $\sqrt{f}\,\omega_0$ [Eq.~\eqref{eq:omega0}], tens to a few hundred
radians per second.
At that frequency the surrounding air's vorticity diffuses outward
from a bead only across an oscillatory boundary layer of thickness
$\delta = \sqrt{2\nu_m/(\sqrt{f}\,\omega_0)}$, with $\nu_m = \eta_m/\rho_m$ the
kinematic viscosity of air \cite{happel2012low}.
Over the \qtyrange{19}{34}{\hertz} trap-frequency range already
quoted above, $\delta \approx \qtyrange{0.4}{0.5}{\milli\meter}$, which is
comparable to the beads' radii, $a/\delta \approx \order{1}$.
The classical unsteady-Stokes result for an oscillating sphere
increases the dissipative drag coefficient over its steady value by
a factor of order $(1 + a/\delta)$ and adds a comparable reactive
(added-mass) contribution.
The corrections could increase the damping ratio proportionately.

\subsection{Hydrodynamic interactions}
\label{sec:approx_hydro}

The total drag on the pair of spheres is modeled as the sum of the
independent Stokes drags on the individual beads, neglecting
hydrodynamic coupling between them.
At contact, where the gap between sphere surfaces vanishes,
lubrication forces and the flow reflected from one sphere onto
the other provide corrections that are not captured by this
superposition approximation.
A more complete treatment would replace the diagonal drag
matrix with the full Oseen-Burgers mobility tensor for
a two-sphere system.

A second, independent approximation concerns each bead's own rotation.
$I=M\ell_a\ell_b$ and $\Gamma_I$ [Eq.~\eqref{eq:GammaI}] treat each
bead as a point mass at its own center, with neither bead spinning
about its own axis as the pair rotates. This requires the bead
surfaces to slide freely past each other at contact, at relative
speed $(a+b)\dot\psi$.

Real bead surfaces carry microscopic asperities that more plausibly
lock each bead's orientation to $\hat n$ than let it slide, so that
the beads co-rotate rigidly instead. In that limit each bead's own
rotational inertia and rotational Stokes drag add to the point-mass
values above:
\begin{align}
  I &\to I + \frac{2}{5}\,M\,\frac{a^5+b^5}{a^3+b^3}, \\
  \Gamma_I &\to \Gamma_I + \frac{4}{3}\,\Gamma\,(a^2-ab+b^2).
\end{align}
At the $(ka,kb)=(\num{0.35},\num{1.5})$ geometry used throughout this
work, these terms raise $I$ by roughly a factor of \num{22} and
$\Gamma_I$ by roughly a factor of \num{5}.

We neglect this correction for the same reason we neglect the
comparable near-field enhancement of $F_\text{nr}$ itself
(\S~\ref{sec:approx_scattering}): both are leading-order estimates,
likely uncertain by a comparable or larger factor at the
close-contact geometries this system requires, so refining one
without the other would not obviously improve, and could worsen,
agreement with experiment. The qualitative picture should be robust
to either choice; the quantitative threshold locations should not be
taken as more precise than an order of magnitude.

Each of these approximations is expected to affect the spinning
threshold and the quantitative values of $r_\text{lc}$ and $\Omega$.
The finite-Reynolds-number and unsteady-drag corrections above raise
the effective damping by less than an order of magnitude; the
rotational correction to $I$ and $\Gamma_I$ can be substantially
larger for asymmetric pairs, and the sign of the single-scattering
correction to $F_\text{nr}$ is not fixed by these arguments alone.
The qualitative picture should be robust to all of these corrections;
the quantitative thresholds should not be taken as more precise than
an order of magnitude.

\section{Numerical methods}
\label{app:numerics}

\subsection{Librational orbit continuation}

Newton shooting on the full nonlinear equations of motion, Eq.~\eqref{eq:eom},
converges reliably only once anchored on a fixed, nonzero value of one
coordinate rather than on a phase condition: without this, the trivial
(non-oscillating) equilibrium is itself a spurious solution of the shooting
equations at any period, and an underconstrained solver can converge onto
it in disguise. Seeded from the linear Hopf eigenvector at small
$\varepsilon$ and continued to finite amplitude by natural continuation,
this reliably traces the librational orbit family.

\subsection{Poincar\'e sections}

Each trajectory presented in Fig.~\ref{fig:chaosonset}(a)
is seeded near the static equilibrium and then evolves through
the librational pathway to its asymptotic dynamical state.
Poincar\'{e} sections are sampled stroboscopically at intervals
of the librational period, $2 \pi / \omega$, and are recorded
through $N = \num{2000}$ periods after initial transients
have damped out.

\subsection{Correctness check}

As a correctness check, the tangent to the family of orbits related by the
system's continuous rotational symmetry is required to be an eigenvector
of $\tensor{M}$ with eigenvalue \num{1} to numerical precision; this holds
to better than one part in \num{e10} throughout the range investigated.

\subsection{Lyapunov exponent estimation}

Reference trajectories are seeded near
the (unstable) static equilibrium, $\psi$ displaced by
\qty{e-3}{\radian} from $\pi$, triggering growth through the
librational instability, and integrated for a settling transient
before the Lyapunov estimate begins:
\num{3000} time units for the reference values shown in
Fig.~\ref{fig:chaosonset}(b), and \num{2000} reference periods for
the examples in Fig.~\ref{fig:chaosonset}.
The densities $p(\lambda\,|\,\varepsilon)\,\omega_0$ in Fig.~\ref{fig:chaosonset}(b) use \num{40}
random initial conditions at each $\varepsilon$:
position uniform in radius, $\num{0.5} \leq kr \leq \num{5}$, with uniform angle,
Gaussian velocities and spin rate with standard deviation \num{0.3},
and uniform orientation $\psi$, each settled for \num{2000} reference periods.
Here, $\lambda$ is estimated over
$\num{5e4} \leq t \leq \num{e5}$ to remove the $1/t$ bias
of the running estimate on tori, and $p(\lambda\,|\,\varepsilon)$ is a
Gaussian kernel density estimate on the figure's
symmetric-logarithmic axis, transformed to the dimensionless
density in $\lambda / \omega_0$, normalized to unity at each
$\varepsilon$, and smoothed in $\varepsilon$.
From the
settled state, a second trajectory is seeded a distance
$d_0 = \num{e-8}$ away, and the pair is propagated together,
renormalizing the separation back to $d_0$ every $\Delta t = \num{20}$
time units (of order one orbital period) while accumulating the
logarithm of the stretch factor; $\lambda(t)$ is this accumulated sum
divided by elapsed time. The result is insensitive to the direction of
the initial perturbation: five independent choices give
$\lambda(t=\num{1.5e4}) \in [\num{1.35}, \num{1.58}] \times 10^{-3}\,\omega_0$,
close to the value quoted above, and
$\lambda(t=\num{3e5}) \in [\num{6.97}, \num{8.10}] \times 10^{-5}\,\omega_0$,
bracketing it.

\bibliography{chasing}

\end{document}